\documentclass{amsart}

\newtheorem{theorem}{Theorem}[section]
\newtheorem{corollary}[theorem]{Corollary}

\theoremstyle{definition}
\newtheorem{definition}[theorem]{Definition}

\theoremstyle{remark}
\newtheorem{remark}[theorem]{Remark}

\numberwithin{equation}{section}

\def\akchoose(#1,#2){\left({#1 \atop #2}\right)}
\def\braket#1{\langle#1\rangle}
\def\ket#1{|#1\rangle}

\def\mod{\,\textup{mod}\,}
\def\Ai{\textup{Ai}}
\def\Bi{\textup{Bi}}
\def\hh[#1,#2,#3]{h_{#3}^{#1,#2}}
\def\bibref[#1]{\cite{#1}} 
\def\KP{\textup{KP}}
\def\C{\mathbb{C}} \def\S{\mathbb{S}} \def\N{\mathbb{N}}
\def\Z{\mathbb{Z}} \def\calL{\mathcal{L}} \def\P{\mathbb{P}}

\begin{document}

\title{Grassmannians, Nonlinear Wave Equations and Generalized Schur Functions}

\author{Alex Kasman}
\address{Mathematical Sciences Research Institute\\
Berkeley, CA 94720}
\email{kasman@msri.org}
\thanks{This research was supported by an MSRI Postdoctoral Fellowship.}


\begin{abstract}
A set of functions is introduced which generalizes the famous Schur
polynomials and their connection to Grasmannian manifolds.  These
functions are shown to provide a new method of constructing solutions
to the KP hierarchy of nonlinear partial differential equations.
Specifically, just as the Schur polynomials are used to expand
tau-functions as a sum, it is shown that it is natural to expand a
\textit{quotient} of tau-functions in terms of these generalized Schur
functions. The coefficients in this expansion are found to be
constrained by the Pl\"ucker relations of a grassmannian.
\end{abstract}

\maketitle

\markboth{Alex Kasman}{Grassmannians, Wave Equations and Generalized Schur Functions}

\section{Introduction}

The \textit{KP hierarchy} of nonlinear partial differential equations
is an important example of the interplay between dynamics and
geometry.  As dynamical systems, the equations
of the KP hierarchy are used to model everything
from ocean waves to elementary particles.  However, the underlying
mathematical structure is extremely geometric, being intimately
related with the moduli of vector bundles over complex projective algebraic
curves, coordinate systems for differential geometric objects, and
especially the geometry of Grassmannian manifolds.

In this paper, a new set of functions is introduced which generalizes
the well known Schur polynomials \bibref[MacD,PS,SW].  Like the Schur
polynomials, these \textit{$N$-Schur functions} are shown to be
relevant to the KP hierarchy and directly related to its underlying
geometric structure.

The remainder of this introduction contains a motivating example
followed by a review of the theory of the KP hierarchy.  Section~2
provides new definitions while Section~3 contains the new results.
The paper concludes with a brief discussion in Section~4.

{\bf Acknowledgements:} I appreciate many helpful discussions with M.
Bergvelt, F. Sottile, G. Wilson and especially E. Previato.  And, of
course, I am grateful to the anonymous referee for helpful comments
and to K. Kuperberg for organizing the \textit{Session on Geometry in
Dynamics} at the 1999 Joint Mathematics meeting in San Antonio, Texas
at which these results were first presented.

\subsection{An Example}\label{sec:examp}

Consider the nonlinear partial differential equation known as the \textit{KP equation}:
\begin{equation}
\KP[u(x,y,t)]:=\frac34 u_{yy}-\left(u_t-\frac14(6uu_x+u_{xxx})\right)_x=0\label{eqn:KPeq}
\end{equation}
This equation was originally derived in \bibref[KP] as a model in
fluid dynamics and contains as a one dimensional reduction (assuming
$u_y=0$) the famous KdV equation \bibref[KdV].   This equation shows
up in many applications, for instance in describing ocean waves
\bibref[Segur].  Moreover, although it is clearly an infinite dimensional
dynamical system, it also contains as appropriate reductions certain classical finite dimensional Hamiltonian
particle systems \bibref[cmbis,KrCM,KrCM2,rothstein,ruij,ShiotaCM,W2].
Forgetting about its many applications, 
this equation may not \textit{appear} to be significantly different than other
nonlinear PDEs.  However, it actually is a remarkably special sort of
equation in that it is \textit{completely integrable}.  This term has
technical meaning, but here it will be used loosely to mean only that we
can write many explicit solutions to this equation and even exactly
solve the initial value problem for certain general classes of initial
conditions (cf.\ \bibref[bd]).


If the equation were \textit{linear}, we would have a geometric
interpretation for the set of solutions as a \textit{vector space}.
However, since \eqref{eqn:KPeq} is nonlinear, we do not initially have
any \textit{geometric} notion of the solution space.  The following
example will demonstrate the sense in which solutions to the KP
equation can similarly have the structure of an algebraic variety.

It is well known that one may construct solutions to the KP equation
from Schur polynomials (cf.~\bibref[SW]).  The main result of this
paper is a generalization of this result.
Without any
explanation, let me provide you with six solutions constructed from
the \textit{generalized} Schur functions to be introduced below.  Define
$$
\tau_0=\exp\left(\frac{-x^3}{6(1+3t)}\right)\qquad \tau_1=1
$$
\begin{eqnarray*}
\tau_2 &=&    y + {\frac{{{\left( 1 + 3t \right) }^{{\frac{1}{3}}}}
       \Gamma({\frac{1}{3}})\Gamma({\frac{2}{3}})}{4^{\frac43}}}
       \left( - {2^{{\frac{1}{3}}}}x
            \left( 3{{\Ai(\theta )}^2} - 
              {{\Bi(\theta )}^2} \right) \right.
\\&&
\left.  + 
         {{\left( 1 + 3t \right) }^{{\frac{1}{3}}}}
          \left( 3{{\Ai'(\theta )}^2} - 
            {{\Bi'(\theta )}^2} \right)  \right)
\end{eqnarray*}
\begin{eqnarray*}
\tau_3 &=& 
\frac{{{\left( 3 + 9t \right) }^{{\frac{1}{3}}}}
      {{\Gamma({\frac{2}{3}})}^2}}{8}
      \left( -6x{{\Ai(\theta )}^2} - 
        4{\sqrt{3}}x\Ai(\theta )
         \Bi(\theta ) \right. \cr && \left.- 
        2x{{\Bi(\theta )}^2} + 
        {4^{{\frac{1}{3}}}}{{\left( 1 + 3t \right) }^{{\frac{1}{3}}}}
         \left( 3{{\Ai'(\theta )}^2} + 
           2{\sqrt{3}}\Ai'(\theta )
            \Bi'(\theta ) + {{\Bi'(\theta )}^2}
            \right)  \right) 
\end{eqnarray*}
\begin{eqnarray*}
\tau_4 &=&  -{\frac{1}{2}} + {\frac{{{\left( 1 + 3t \right) }^{{\frac{1}{3}}}}
       {{\Gamma({\frac{1}{3}})}^2}}       {8({3^{{\frac{5}{6}}}})}}
       \left( -3({2^{{\frac{1}{3}}}}){\sqrt{3}}x
          {{\Ai(\theta )}^2} + 
         6({2^{{\frac{1}{3}}}})x\Ai(\theta )
          \Bi(\theta )\right.
\\ &&
\left. - 
         {2^{{\frac{1}{3}}}}{\sqrt{3}}x
          {{\Bi(\theta )}^2} + 
         {{\left( 1 + 3t \right) }^{{\frac{1}{3}}}}
          \left( 3{\sqrt{3}}{{\Ai'(\theta )}^2} - 
            6\Ai'(\theta )\Bi'(\theta ) + 
            {\sqrt{3}}{{\Bi'(\theta )}^2} \right)  \right)
\end{eqnarray*}
\begin{eqnarray*}
\tau_5 &=&    -y + {\frac{{{\left( 1 + 3t \right) }^{{\frac{1}{3}}}}
       \Gamma({\frac{1}{3}})\Gamma({\frac{2}{3}})}{4^{\frac43}}}
       \left( - {2^{{\frac{1}{3}}}}x
            \left( 3{{\Ai(\theta )}^2} - 
              {{\Bi(\theta )}^2} \right) \right.
\\&&\left.
  + 
         {{\left( 1 + 3t \right) }^{{\frac{1}{3}}}}
          \left( 3{{\Ai'(\theta )}^2} - 
            {{\Bi'(\theta )}^2} \right)  \right) 
\end{eqnarray*}
\begin{eqnarray*}
\tau_6 &=& 
  {\frac{-x \left( 1 + 3t \right)  }{2}} + {y^2} + 
   {\frac{{{\left( 3 + 9t \right) }^{{\frac{1}{3}}}}
       {{\Gamma({\frac{2}{3}})}^2}}{16}}
       \left( 6x{{\Ai(\theta )}^2} + 
         4{\sqrt{3}}x\Ai(\theta )
          \Bi(\theta )  \right.
\\ &&
\left.
+         2x{{\Bi(\theta )}^2} - 
         {2^{{\frac{2}{3}}}}{{\left( 1 + 3t \right) }^{{\frac{1}{3}}}}
          \left( 3{{\Ai'(\theta )}^2} + 
            2{\sqrt{3}}\Ai'(\theta )
             \Bi'(\theta ) + {{\Bi'(\theta )}^2}
             \right)  \right) 
\end{eqnarray*}
and
$$
\theta=  {\frac{{2^{{\frac{1}{3}}}}x}
    {{{\left( 1 + 3t \right) }^{{\frac{1}{3}}}}}}
$$
Here Ai and Bi are the standard Airy functions satisfying the
differential equation $f''(x)=xf(x)$ and having Wronskian determinant
$2/(\sqrt{3}\Gamma({1/3})\Gamma({2/3}))$.  Then note that each
function $$u_{i}(x,y,t)=2\frac{\partial^2}{\partial x^2}
\log(\tau_0(x,t)*\tau_{i}(x,y,t))\qquad (1\leq i \leq 6)$$ is a
solution to the KP equation\footnote{That the function
$u_1=-2x/(3t+1)$ is a solution is simple enough to check by hand.  I
might suggest using a computer to verify the other solutions.}.  The
point here is not to understand \textit{how} these solutions were
found (this will be explained later), but rather to see the way in
which they demonstrate the algebro-geometric structure underlying the
equation.

Consider an arbitrary linear combination of these \textit{tau-functions}:
$$
\tau(x,y,t):=\tau_{0}\cdot\sum_{i=1}^6\pi_i\tau_i\qquad\pi_{i}\in\C
$$
and the corresponding function $u(x,y,t)=2\partial^2/\partial x^2
(\log\tau)$.  We know that this function is a solution to \eqref{eqn:KPeq}
for \textit{certain} choices of the coefficients $\pi_{i}$; in particular all
but one could be equal to zero.  \textit{Do any other 
combinations lead to a KP solution? }

If every such combination gave a solution, then this would be a
six-dimensional vector space of solutions.  This is certainly not the
case since, for instance, one \textit{cannot} have all $\pi_i=0$.  Let
us naively answer the question by simply inserting\footnote{Again, use
a symbolic algebra computer program such as
\textit{Mathematica}. Actually, as the experts know, it is easier to
use the bilinear form (cf.~\bibref[Jimbo]):
$$
 \frac34\tau\tau_{yy}   -\frac34{\tau_{y}^2}+
   \tau_x\tau_t -
   \tau\tau_{xt}
+ \frac34{\tau_{xx}^2}
 -
   \tau_{x}\tau_{xxx} +
   \frac14\tau\tau_{xxxx}=0
$$
of the KP equation rather
than \eqref{eqn:KPeq}.} this function $u(x,y,t)$ into Equation
\eqref{eqn:KPeq}.  One finds, after algebraic simplification, that
$\KP[u]=0$ is nothing but the homogeneous algebraic equation $$
\pi_1\pi_6-\pi_3\pi_5+\pi_2\pi_4=0.
$$
This
particular algebraic equation describes a well known projective geometric
object: the Grassmannian $Gr_{2,4}$ \bibref[HodgePedoe].

The rest of this paper will attempt to explain this example, describe
the way in which one can do the same to obtain the equations for any
finite dimensional Grassmannian, and especially to emphasize its
connection to the $N$-Schur functions.

\subsection{The KP Hierarchy}\label{sec:KP}

To understand the structure underlying the
integrability of \eqref{eqn:KPeq}, it is convenient to
consider the KP equation as only part of an infinite hierarchy of
equations, and to consider the variables $x$, $y$ and $t$ as only the
first three variables in a hierarchy of infinitely many variables.

The KP hierarchy is an infinite set of compatible dynamical systems on
the space of monic pseudo-differential operators of order one.  A
pseudo-differential operator is a Laurent series in the symbol
$\partial$ with coefficients that are functions of the variable $x$.
The multiplication of these operators is defined by the relationships
$$
\partial\circ f(x)=f(x)\partial+f'(x)\qquad \partial^{-1}\circ
f(x)=f(x)\partial^{-1}-f'(x)\partial^{-2}+f''\partial^{-3}+\cdots
$$
In other words, $\partial=d/dx$ and $\partial^{-1}$ is
its formal inverse.  Contained within the ring of pseudo-differential
operators is the ring of ordinary differential operators, those having
only non-negative powers of $\partial$.

An \textit{initial condition} for the KP hierarchy is any
pseudo-differential operator of the form
\begin{equation}
\calL=\partial + w_1(x)\partial^{-1}+ w_2(x)\partial^{-2}+\cdots.\label{eqn:initcond}
\end{equation}
The KP hierarchy is the set of dynamical systems defined by the
evolution equations
\begin{equation}
\frac{\partial}{\partial t_i}\calL=[\calL,(\calL^i)_+]\qquad i=1,2,3,\ldots\label{eqn:KPhier}
\end{equation}
where the ``$+$'' subscript indicates projection onto the differential
operators by simply eliminating all negative powers of $\partial$ and
$[A,B]=A\circ B- B\circ A$.  In fact, since all of these flows commute
(for $i=1,2,\ldots$) one can think of a solution of the KP hierarchy
as a pseudo-differential operator of the form
\eqref{eqn:initcond} whose coefficients depend on the time variables
$t_1, t_2,\ldots$ so as to satisfy \eqref{eqn:KPhier}.  (Note also
that the first equation, $i=1$, leads to the conclusion that $t_1=x$
so these names will be used interchangeably.  Similarly it is common
to use $y=t_2$ and $t=t_3$.)

\subsubsection{The Tau-Function}

Remarkably, there exists a convenient way to encode all information
about the KP solution $\calL$ in a single function of the time
variables $t_1,t_2,\ldots$.  Specifically, each of the coefficients
$w_i$ of $\calL$ can be written as a rational function of this
function $\tau(t_1,t_2,\ldots)$ and its derivatives \bibref[SW].
Alternatively, one can construct $\calL$ from $\tau$ by letting $W$ be
the pseudo-differential operator
$$
W=\frac{1}{\tau}
\tau(t_1-\partial^{-1},t_2-\frac12\partial^{-2},\ldots)
$$
and then $\calL:=W\circ\partial\circ W^{-1}$ is a solution to the KP
hierarchy \bibref[AvM].  

Every solution to the KP hierarchy can be written this way in terms of
a tau-function, though the choice of tau-function is not unique.  For
example, note that one may always multiply $W$ on the right by any constant
coefficient series $1+O(\partial^{-1})$ without affecting the corresponding
solution.  More significant to the present paper is the elementary
observation that multiplying the tau-function by any constant will not
change the associated solution.  So, it is reasonable to consider two
tau-functions to be equivalent if they differ by multiplication of a
constant (independent of $\{t_i\}$).  Then, like the Pl\"ucker
coordinates of a Grassmannian \bibref[HodgePedoe], the tau-function is uniquely defined
only up to this \textit{projective} equivalence.

\subsection{Significance of the KP Hierarchy}
If $\calL$ is a solution to the KP hierarchy then the function
$$u(x,y,t)=-2\frac{\partial}{\partial
x}w_1(x,y,t,\ldots)=2\frac{\partial^2}{\partial x^2}\log \tau$$ is a
solution of the KP equation \eqref{eqn:KPeq}.  Moreover, many of the
other equations that show up as particular reductions of the KP
hierarchy have also been previously studied as physically relevant
wave equations.  The KP hierarchy also arises in \textit{string
theories} of quantum gravity \bibref[vM], the probability
distributions of the eigenvalues of random matrices
\bibref[newAvM,Tracy], and the description of coordinate systems in
differential geometry \bibref[doliwa].

Certainly one of the most significant comments which can be made
regarding these equations, which is a consequence of the
form \eqref{eqn:KPhier}, is that all of these equations are completely
integrable.  Among the many ways to solve the equations of the KP
hierarchy are several with connections to the algebraic geometry of
``spectral curves'' \bibref[Itsbook,Kr,PW,Pr,SW].
(Conversely, through this same correspondence the KP hierarchy
provides an answer to the famous Schottky problem in algebraic
geometry \bibref[Mul,Shiota], representing \textit{another} direction
to the interaction between dynamics and geometry.)  However, more
relevant to the subject of this note is the observation of M. Sato
that the geometry of an \textit{infinite dimensional Grassmannian}
underlies the solutions to the KP hierarchy \bibref[Sato].

\subsubsection{$N$-KdV and the Vector Baker-Akhiezer function}

Of particular interest below are the solutions $\calL$
of the KP hierarchy that have the property that 
$L=\calL^N=(\calL^N)_+$ is an \textit{ordinary} differential operator.
We say that solutions of the KP hierarchy with this property are
solutions of the of the $N$-KdV hierarchy.  One can easily check from
\eqref{eqn:KPhier} that these solutions are stationary under all flows
$t_i$ where $i\equiv0\mod N$.  For instance, the solutions of the $2$-KdV
hierarchy are independent of all even indexed time parameters and
consequently give solutions $u$ to
\eqref{eqn:KPeq} which are independent of $y$ and therefore solve the
KdV equation.

Associated to any chosen solution $L=\calL^N$ of the $N$-KdV hierarchy, we
 associate an $N$-vector valued function of the variables
$\{t_i\}$ and the new \textit{spectral parameter} $z$.  
Following \bibref[PW] we define the \textit{vector Baker-Akhiezer}
function to be the unique function $\vec\psi(z,t_1,t_2,\ldots)$
satisfying 
\begin{equation} L\vec\psi=z\vec\psi\qquad \frac{\partial}{\partial
t_i}\vec\psi=(\calL^i)_+ \vec\psi \label{eqn:psivec}
\end{equation}
and such that the $N\times N$
Wronskian matrix 
\begin{equation}
\Psi(z,t_1,t_2,\ldots):=\left(\begin{matrix}
\vec\psi\cr
\frac{\partial}{\partial x} \vec\psi
\cr
\vdots\cr
\frac{\partial^{N-1}}{\partial x^{N-1}} \vec\psi\end{matrix}\right)
\label{eqn:Psi}
\end{equation}
is the identify matrix when evaluated at
$0=t_1=t_2=\ldots$.  (We will ignore here the case in which the matrix
is undefined at this point.  In fact, this is not a serious problem
since such singularities are isolated and thus the problem can be
resolved by using the KP flows.)

\section{A Generalization of the Schur Functions}\label{sec:nschur}

Let $N\in\N$ and consider the variables $\hh[i,j,k]$ ($1\leq i,j\leq
N$, $k\in\N$) which can be conveniently grouped into $N\times N$
matrices
$$
H_k=\left(\begin{matrix}
\hh[1,1,k]& \cdots& \hh[1,N,k]\cr
\vdots&\ddots&\vdots\cr
\hh[N,1,k]& \cdots& \hh[N,N,k]
\end{matrix}\right)\qquad k=0,1,2,3,4,\ldots
$$
Moreover, these matrices will be grouped into the infinite matrix
$M_{\infty}$ 
$$
M_{\infty}=\left(\begin{matrix}
\vdots&\vdots & \vdots& \cdots \cr
H_2&H_3&H_4&H_5&\cdots&\cdots\cr
H_1&H_2&H_3&H_4&\cdots&\cdots\cr
H_0&H_1&H_2&H_3&\cdots&\cdots  \hbox to 0pt{\hskip 1pc
$\}$\hbox{rows 0 through $N-1$}}\cr
0&H_0&H_1&H_2&H_3&\cdots\hbox to 0pt{\hskip 1pc
$\}$\hbox{rows $N$ through $2N-1$}}\cr
0& 0&H_0&H_1&H_2&\cdots\cr
0& 0& 0&H_0&H_1&\cdots\cr
\vdots&&&&\ddots\end{matrix}\right)\qquad\qquad
$$
It is convenient to label the rows of this matrix by the integers with
$0$ being the first row with $H_0$ at the left and increasing downwards.

We wish to define a set of functions in these variables indexed by
partitions of integers.  Specifically, the index set $\S$ will be the
set of increasing sequences of integers whose values are eventually
equal to their indices
$$
\S=\{(s_0,s_1,s_2,\ldots)\ |\ s_{j+1}>s_j\in\Z\ \textup{and}\ \exists
m\ \textup{such\ that}\ j=s_j\ \forall j>m\}.
$$
Of particular interest here will be
the special case $0:=(0,1,2,3,\ldots)\in\S$. 

\begin{definition} For any $S\in\S$ let $f_S^N:=\det(M_S\cdot M_0^{-1})$
where the infinite matrix $M_S$ is the matrix whose $j^{th}$ row is
the $s_j^{th}$ row of $M_{\infty}$. Then we have, for instance, that
$$
M_0=\left(\begin{matrix}
H_0&H_1&H_2&H_3&\cdots&\cdots\cr
0&H_0&H_1&H_2&H_3&\cdots\cr
0& 0&H_0&H_1&H_2&\cdots\cr
0& 0& 0&H_0&H_1&\cdots\cr
\vdots&&&&\ddots\end{matrix}\right).
$$
Equivalently, one could say that the element in 
position $(l,m)$ ($0\leq l,m\leq \infty$) of the matrix $M_S$ is given
by
$$
(M_S)_{l,m}=\hh[i,j,k]\qquad i=
1+(l\mod N),\ j=1+(s_m\mod N),\ k={\left\lfloor \frac
lN\right\rfloor-\left\lfloor \frac{s_m}N\right\rfloor}
$$
where $\hh[i,j,k]=0$ if $k<0$.
\end{definition}
\begin{remark} The definition of $M_S$ should remind one of the definition of
the Pl\"ucker coordinates \bibref[HodgePedoe].  This is no coincidence; in fact we
will see below that the $N$-Schur functions arise naturally in the
context of an infinite dimensional Grassmannian.
\end{remark}
\begin{remark} You do not have to be comfortable with infinite matrices to
work with the $N$-Schur functions.
Note that given any $m\in\N$ such that $s_i=i$ for all $i>mN$,
the matrix $M_S\cdot M_0^{-1}$ looks like the identity matrix below
the $mN^{th}$ row.  Consequently, the easiest way to actually compute
these functions is as two finite determinants
$$
f_S^N=\frac{\det(M_S|_{mN\times mN})}{(\det H_0)^m}
$$
where $M_S|_{mN\times mN}$ denotes the top left block of size
$mN\times mN$ of the matrix $M_S$.  For example,
regardless of $N$, $f_0^N\equiv 1$.  However, for
$S=(-2,1,2,3,\ldots)$ one finds instead 
$$
f_S^1=\frac{\hh[1,1,2]}{\hh[1,1,0]}\qquad
f_S^2=\frac{\hh[1,1,1]\hh[2,2,0]-\hh[1,2,0]\hh[2,1,1]}{\hh[1,1,0]\hh[2,2,0]-\hh[1,2,0]\hh[2,1,0]}.
$$
\end{remark}
\begin{remark} In fact, in the case $N=1$ and
$\hh[1,1,0]=1$, the functions $\{f_S^1\}$ are the famous Schur polynomials
\bibref[MacD] (cf.\ \bibref[PS,SW]).
Similarly, if we assume in general that
$\det H_0=1$, then
all $f_S^N$ are polynomials.   
\end{remark}

\begin{remark} If we consider $\hh[i,j,k]$ to have weight $kN+i-j$ then the
function $f_S^N$ is homogeneous of weight $\sum_{j=0}^{\infty}s_j-j$.
\end{remark}

\section{Tau-Functions, Determinants on an Infinite Grassmannian and $N$-Schur Functions}\label{sec:mrs}

\textit{Quotients} of tau-functions have recently played a prominent
role in several papers on bispectrality \bibref[BHYbisp,KR], Darboux
transformations \bibref[BHYdt,aam] and random matrices
\bibref[newAvM].  In \bibref[aam] these quotients themselves are
computed as a determinant of the action of a matrix valued function on
the frame bundle of the grassmannian $Gr^N=Gr(H^N)$ (cf.\
\bibref[PW]). Here we will see that such a determinant can be
 decomposed into a sum of $N$-Schur functions when given
appropriate dependence on the time variables of the KP hierarchy.

\subsection{The Grassmannian $Gr^N$}

Let us recall notation and some basic facts about 
infinite dimensional grassmannians.  Please refer to
\bibref[aam,PS,PW,SW] for additional details.

Let $H^N=L^2(S^1,\C)$ be the Hilbert space of square-integrable vector
valued functions $S^1\to\C^N$, where $S^1\subset\C$ is the unit
circle.  Denote by $e_i$ ($0\leq i \leq N-1$) the $N$-vector which has
the value 1 in the $i+1^{st}$ component and zero in the others.  We
fix as a basis for $H^N$ the set $\{e_i|i\in\Z\}$ with
$$
e_i:=z^{\lfloor \frac i n \rfloor}e_{(i\mod N)}.
$$
The Hilbert space has the decomposition
\begin{equation}
H^N=H_+^N\oplus H_-^N\label{eqn:split}
\end{equation}
where these subspaces are spanned by the basis elements with
non-negative and negative indices respectively.
Then $Gr^N$ denotes the grassmannian of all closed subspaces $W\subset
H^N$ such that the orthogonal projection $W\to H^N_-$ is a compact
operator and such that the orthogonal projection $W\to H_+^N$ is
Fredholm of index zero \bibref[PS,PW].

Associate to any basis $\{w_0,w_1,\ldots\}$ for a point $W\in Gr^N$ 
the linear map $w$
\begin{eqnarray*}
w:H_+^N&\to&W\\
e_i&\mapsto&w_i.
\end{eqnarray*}
The basis is said to be \textit{admissible} if
$w$ differs from the identity by an element of trace class
\bibref[Simon].  The \textit{frame bundle} of $Gr^N$ is the set of
pairs $(W,w)$ where $W\in Gr^N$ and $w:H_+^N\to W$ is an admissible
basis.

There is a convenient way to embed $Gr^N$ in a projective space.
Let $\Lambda$ denote the infinite alternating exterior algebra
generated by the alternating tensors
$$
\{e_{s_0}\wedge e_{s_1}\wedge
e_{s_2}\wedge\cdots|(s_0,s_1,s_2,\ldots)\in\S\}.
$$
To any point $(W,w)$ in the frame bundle we associate the alternating
tensor
$$
\ket{w}:=w_0\wedge w_1\wedge w_2\wedge\cdots\in\Lambda.
$$
Note in particular that $\ket{\cdot}$ is projectively well defined on
the entire fiber of $W$ (i.e.\ for two admissible bases of $W$ we have
$\ket{w}=\lambda\ket{w'}$ for some non-zero
constant $\lambda$).  Consequently, $\ket{W}$ is a well defined
element of the projective space $\P\Lambda$.  

The Pl\"ucker coordinates of $W$ are
the coefficients $\braket{S|W}$ in the unique expansion
$$
\ket{W}=\sum_{S\in\S} \braket{S|W} e_{s_0}\wedge e_{s_1}\wedge
e_{s_2}\wedge \cdots
$$
and are therefore well defined as a set up to a common multiple.
Alternatively, given an admissible basis $w$ for $W$, $\braket{S|W}$
is the determinant of the infinite matrix made of the rows of $w$
indexed by the elements of $S$. 

\subsection{Main Results}

Let $g$ be an $N\times N$ matrix valued function of $z$ with expansion
\begin{equation}
g=\sum_{k=0}^{\infty} H_k z^k\qquad H_k=\left(\begin{matrix}
\hh[1,1,k]& \cdots& \hh[1,N,k]\cr
\vdots&\ddots&\vdots\cr
\hh[N,1,k]& \cdots& \hh[N,N,k]
\end{matrix}\right)\label{eqn:gform}
\end{equation}
such that an inverse matrix $g^{-1}$ exists for all $z$.  We will view
$g\in GL(H^N)$ as an operator on $Gr^N$ and demonstrate that the
$N$-Schur functions arise naturally in this context.

In general, an operator on the frame bundle \bibref[PS] is a pair
$A=(g,q)$ where $g\in GL(H^N)$ with the form 
\begin{equation}
g=\left(\begin{matrix}a&b\cr
c&d\end{matrix}\right)
\end{equation}
relative to the splitting \eqref{eqn:split} and $q:H_+^N\to H_+^N$
such that $a\cdot q^{-1}$ differs from the identity by an operator of
trace class.  The action is given by
$$
A:(W,w)\mapsto (gW,gwq^{-1}).
$$

In the particular case \eqref{eqn:gform} of interest here 
$c=0$ and we simply let $q=a$ so that $aq^{-1}$ is
the identity matrix.  We will write $\ket{g|w}=\ket{gwa^{-1}}$ for the
action of $g$ on the frame bundle.  Moreover, since this action is
well defined on projective equivalence classes we will write
$\ket{g|W}$ for the class containing $\ket{g|w}$ with any admissible
basis $w$ of $W$.

A main result of this paper is then the observation that for any
point $W\in Gr^N$ the determinant $\braket{0|g|W}$ can be written in
terms of the Pl\"ucker coordinates of $W$ and the $N$-Schur functions:
\begin{theorem}\label{thm:theorem}
For $W\in Gr^N$ and $g$ as in \eqref{eqn:gform}
$$
\braket{0|g|W}=\sum_{S\in\S}\braket{S|W} f_S^N.
$$
\end{theorem}

\begin{proof}
The proof is elementary in the case $W=W_S$ with basis
$\{e_{s_0},e_{s_1},e_{s_2},\ldots\}$.  In fact, this is essentially
the definition of $f_S^N$ since the matrix representation of the
operator $g$ is precisely $M_{\infty}$ and similarly $q$ is $M_0$.
So, $gwq^{-1}$ is the matrix $M_SM_0^{-1}$ whose determinant is
$f_S^N$. The general case follows from the observation that
multilinearity of determinants is equivalent to the linearity of the
map $\langle 0|g|:\Lambda\to\C$ and expanding $\ket{W}$ as a sum.
\end{proof}

This general result is especially interesting in the case that the
variables $\hh[i,j,k]$ are evaluated as special functions of the KP times
$t_i$ associated to the choice of a solution of the $N$-KdV
hierarchy.  Specifically, associated to the choice of a solution
$\calL$ of the $N$-KdV hierarchy define
\begin{equation}
\hh[i,j,k]:=\frac{1}{k!}\frac{\partial^k}{\partial
z^k}(\Psi^{-1})_{ij}\bigg|_{z=0}\label{eqn:derivform}
\end{equation}
where $\Psi$ is the corresponding Wronskian matrix.
In that case each $N$-Schur function
is a quotient of KP tau-functions:
\begin{theorem}\label{cor}
Let $\calL$ be a solution of the $N$-KdV hierarchy with corresponding
tau-function $\tau_0$ and corresponding matrix $\Psi$ given in \eqref{eqn:psivec} and \eqref{eqn:Psi}.
Give  the $N$-Schur functions dependence on the time variables of the
KP hierarchy through \eqref{eqn:derivform} so that
$$
\Psi^{-1}=\sum_{k=0}^{\infty} H_k z^k.
$$
Then there exists a tau-function $\tau_S$ of the KP
hierarchy so that
$$
f_S^N(t_1,t_2,\ldots)=\frac{\tau_S}{\tau_0}
$$
for every 
 $S\in\S$.  Moreover, it follows that
$$
\tau_0\cdot(\sum_{S\in\S} \pi_S f_S^N)
$$
is a tau-function of the KP hierarchy whenever $\pi_S$ are the
Pl\"ucker coordinates of some point in $Gr^N$.
\end{theorem}
\begin{proof}
Defining $g=\Psi^{-1}\in GL(H^n)$ where $\Psi$ is the matrix
\eqref{eqn:Psi} above, the determinant $\braket{0|g|W}$ is a
(projective) function of the variables $t_i$.  It is shown in
\bibref[aam] (Definition 7.4 and Claim 7.12) that these functions are
quotients of KP tau-functions with a tau-function corresponding to
$\calL$ in the denominator.  Consequently, using the theorem above we
may write these quotients in terms of the $N$-Schur functions to prove
the claim.
\end{proof}

This is, of course, a generalization of the well known result relating
Schur polynomials and the KP hierarchy.  In particular, that result is
the special case of the 1-KdV solution $L=\partial$ for which
$$
\Psi=\exp(\sum t_iz^i).
$$
In that case, of course, the time dependent polynomials in the
variables $h^{i,j}_k$ are also polynomial in the variables $t_i$.  In
general, that will not be the case.

\subsection{Finite Dimensional Grassmannians}

This would then be a good time to describe the construction of
the solutions to the KP equation given in Section~\ref{sec:examp}.
Since that example concerned only the KP equation (and not all of
the equations of the hierarchy) we need only consider the first three
time variables $t_1=x$, $t_2=y$ and $t_3=t$.  
One well known but surprisingly complicated\footnote{I say that this
solution is surprisingly complicated because it does not come from any
of the usual methods of solution.  This solution is \textit{not}
related to a flow on a Jacobian variety of a spectral curve, since it
is ``rank 2''.  This solution is not solvable by the inverse scattering
method since it certainly does not vanish for $x\to\infty$.  This
solution is not even among the many analytically determined solutions
in \bibref[SW].  Moreover, it is this solution which was related to
intersection numbers on an algebro-geometric moduli space by a
conjecture of Witten and theorem of Kontsevich leading to a Fields'
Medal for the latter.} solution to the $2$-KdV
hierarchy is $L_0=\calL^2=\partial^2-2x/(3t+1)$.  It corresponds to the
tau-function $\tau_0$ in the example.  The other functions $\tau_i$
are just $2$-Schur functions given time dependence by
\begin{eqnarray*}
\Psi^{-1} &=&     \phi(y,z)*\left(   \begin{matrix} {\frac{- \Ai'(\Theta ) 
          \Bi(\zeta)   + 
       \Ai(\zeta) 
        \Bi'(\Theta )}{2 
       {{\left( 1 + 3 t \right) }^{{\frac{1}{6}}}}}} & 
   {\frac{{{\left( 1 + 3 t \right) }^{{\frac{1}{6}}}} 
       \left( - \Ai(\zeta) 
            \Bi(\Theta )   + 
         \Ai(\Theta ) 
          \Bi(\zeta) \right) }{2 
       {2^{{\frac{1}{3}}}}}} \cr 
   {\frac{\Ai'(\zeta) 
        \Bi'(\Theta ) - 
       \Ai'(\Theta ) 
        \Bi'(\zeta)}{
       {2^{{\frac{2}{3}}}} {{\left( 1 + 3 t \right) }^{{\frac{1}{6}}}}}} & 
   {\frac{- {{\left( 1 + 3 t \right) }^{{\frac{1}{6}}}} 
         \left( \Ai'(\zeta) 
            \Bi(\Theta ) - 
           \Ai(\Theta ) 
            \Bi'(\zeta) \right) 
           }{2}} \end{matrix}
\right)\\ &=& 
\sum_{k=0}^{\infty} H_kz^k
\end{eqnarray*}
with $\phi(y,z)=\sqrt{3} \Gamma({\frac{1}{3}})
\Gamma({\frac{2}{3}}) e^{-yz}$, $\zeta=4^{-\frac{1}{3}}z$ and $\Theta={\frac{3 {t z} + 2 x + z} {{2^{{\frac{2}{3}}}} {{\left( 1
+ 3 t \right) }^{{\frac{1}{3}}}}}}$.
Then considering Theorem~\ref{cor} with the additional restriction
that all but these six coordinates must be zero gives exactly the
Pl\"ucker relation for $Gr_{2,4}$.  More generally:

\begin{definition}
Let $k<n\in\N$ be positive integers and define $\S_{k,n}\subset\S$ as
$$
\S_{k,n}=\{S\in\S\ |\ k-n\leq s_i \leq k-1\ (0\leq i\leq k-1)\}.
$$
Note that $\S_{k,n}$ contains exactly $\akchoose(n, k)$ elements.
Specifically, every element of $\S_{k,n}$ corresponds to a choice of
$k$ integers between $k-n$ and $k-1$.
Let
$$
\gamma_{k,n}(t_1,t_2,\ldots):=\sum_{S\in \S_{k,n}} \pi_S
f_S^N(t_1,t_2,\ldots)
$$
where $\pi_S\in\C$ are arbitrary parameters and define
$\tau_{k,n}:=\tau_0\cdot\gamma_{k,n}$.  
\end{definition}

We are naturally led to consider the coefficients $\{\pi_S\}$ as
points in the projective space $\P^{m-1}\C$ ($m=\akchoose(n, k)$) and to ask:
\textit{For what points in this projective space is
$\tau_{k,n}$ a tau-function?}  
The answer provided by
Theorem~\ref{cor} above is simply:
\begin{corollary}\label{cor2}
The function $\tau_{k,n}(t_1,t_2,\ldots)$ depending on the
$\akchoose(n, k)$ parameters $\pi_S$ is a KP tau function precisely
when they satisfy the Pl\"ucker relations for the Grassmannian $Gr_{k,n}$.
\end{corollary}
So, this gives us a procedure for deriving the algebraic Pl\"ucker
relations from the differential equations of the KP hierarchy as in
the example.

\section{Discussion}

That the (1-)Schur polynomials give
solutions to the KP hierarchy is often cited as having been the
\textit{clue} which led Sato to recognize the connection between KP
and Grassmannians \bibref[Sato].  This well known case is especially
nice since it comes from $N$-Schur functions associated to the
simplest solution $\tau=1$ and it is used to expand the tau-function
as a sum.  However, the more
general situation discussed here has similar applications in expanding
\textit{quotients} of tau-functions.  (The Schur polynomials
correspond to the special case in which the tau-function in the
denominator is equal to 1, so the tau-quotient is itself also a
tau-function.)  Such tau-quotients associated with Darboux
transformations are themselves of interest \bibref[newAvM,aam]
and the introduction of the $N$-Schur functions provides a means to
expand these as sums with coefficients constrained by algebraic
equations.

One thing which is not yet clear, at least to me, is whether the
$N$-Schur functions have any group theoretic significance in the case
$N>1$.  Certainly, in the case $N=1$ there are very satisfying
explanations of the role of the Schur polynomials in the KP Hierarchy
in terms of their orthogonality \bibref[Sato], highest-weight
conditions \bibref[Jimbo,Engberg] or representations of Heisenberg
algebras \bibref[bergv].  It would be interesting if one could
generalize these algebraic interpretations of the tau-functions of the
KP hierarchy to the situation discussed above.

\end{document}